\documentclass[12pt]{article}
\pdfoutput=1

\usepackage{graphicx}
\usepackage{epsf}

\usepackage{amssymb,graphicx}
\usepackage{hyperref}
\usepackage[T1]{fontenc}
\usepackage[charter]{mathdesign}
\usepackage[font=small,labelfont=bf]{caption}
\usepackage{tabularx}

\usepackage{longtable}
\usepackage{verbatim}
\usepackage{cite}

\usepackage[usenames,dvipsnames]{xcolor}
\definecolor{burgundy}{rgb}{0.565,0.0,0.125}

\usepackage{afterpage}

\def\be{\begin{equation}}
\def\ee{\end{equation}}
\def\bea{\begin{eqnarray}}
\def\eea{\end{eqnarray}}
\def\nn{\nonumber}

\def\d{\dagger}

\def\dot{\cdot}

\def\sL{\mathcal{L}}

\newcommand{\beq}{\begin{eqnarray}}
\newcommand{\eeq}{\end{eqnarray}}
\def\OMIT#1{{}}
\newcommand{\lsim}{\mathrel{\rlap{\lower4pt\hbox{\hskip1pt$\sim$}}
    \raise1pt\hbox{$<$}}}         
\newcommand{\gsim}{\mathrel{\rlap{\lower4pt\hbox{\hskip1pt$\sim$}}
    \raise1pt\hbox{$>$}}}         

\usepackage{titlesec}

\titleformat{\section}
{\color{gray}\normalfont\Large\bfseries}
{\color{gray}\thesection}{1em}{}

\titleformat{\subsection}
{\color{gray}\normalfont\large\bfseries}
{\color{gray}\thesubsection}{1em}{}

\title{\bf \color{gray} Understanding diboson anomalies}

\author{Aqil Sajjad\\
{\small \color{gray} \texttt{sajjad@physics.harvard.edu}}\\
{\em Department of Physics, Harvard University, Cambridge, MA 02138, USA}}

\begin{document}
\maketitle

\begin{abstract}
\noindent 
We conduct a model-independent effective theory analysis of hypercharged fields with various spin structures towards understanding the diboson excess found in LHC run I, as well as possible future anomalies involving $WZ$ and $WH$ modes. Within the assumption of no additional physics beyond the standard model up to the scale of the possible diboson resonance, we show that a hypercharged scalar and a spin 2 particle do not have tree-level $WZ$ and $WH$ decay channels up to dimension 5 operators, and cannot therefore account for the anomaly, whereas a hypercharged vector is a viable candidate provided we also introduce a $Z'$ in order to satisfy electroweak precision constraints. We calculate bounds on the $Z'$ mass consistent with the Atlas/CMS diboson signals as well as electroweak precision data, taking into account both LHC run I and II data.
 \end{abstract}

\section{Introduction}

The Atlas and CMS collaborations have recently reported several excesses in the diboson decay channels with a possible resonance around $2~{\rm TeV}$ in run 1 of the LHC \cite{Aad:2015owa, Khachatryan:2014hpa, CMS:2015gla}. 
The excesses include the $WZ$, $WW$ and $ZZ$ channels with local significances of $3.4\sigma$, $2.6\sigma$ and $2.9\sigma$, respectively with a resonance around $2~{\rm TeV}$ reported by Atlas 
and the $WH$ mode with a resonance around $1.8-1.9~{\rm TeV}$ with a deviation of $2.2\sigma$ according to CMS. The recently announced run 2 results on the other hand do not show any such excess but the data is not enough to rule out the effect at 95$\%$ confidence level \cite{ATLAS-CONF-2015-073, CMS:2015nmz}. Specifically, the luminosities for the $8~{\rm TeV}$ run were $20.3~{\rm fb^{-1}}$ and $20~{\rm fb^{-1}}$ for Atlas and CMS, respectively, whereas those for the $13~{\rm TeV}$ data released in December were $3.2~{\rm fb^{-1}}$ and $2.6~{\rm fb^{-1}}$ for the two collaborations. Consequently, while the run II results put more stringent bounds on the possible $2~{\rm TeV}$ resonances, more data is needed to come to a definite conclusion about the excesses reported in run I. The tightest bound on the cross-section times branching ratio for the $WZ$ channel from LHC run I comes from the $WH$ data through the Goldstone equivalence theorem \cite{Hisano:2015gna}, which gives a 95$\%$ confidence upper limit of about $7~{\rm fb}$ for a $2~{\rm TeV}$ resonance for a $8~{\rm TeV}$ $PP$ collision \cite{Khachatryan:2015bma}. This corresponds to about $54.7~{\rm fb}$ for a $13~{\rm TeV}$ experiment. In run II, the strictest constraint comes from the $WZ$ data, giving an upper bound of about $40~{\rm fb}$ for a $13~{\rm TeV}$ center of mass energy, which is smaller but still large enough to leave open the possibility that further data could lead to a discovery of a new particle.

Beyond the $2~{\rm TeV}$ LHC run I diboson excess, the $WZ$ channel could also potentially arise in future experiments at other energies and will therefore also be an important part of future searches for new physics. In this backdrop, it is worth developing a framework for understanding such diboson excesses. The purpose of this paper is to offer a simple model-independent effective theory perspective for understanding charged resonances with diboson decays. The motivation for focusing on charged particles is partly that the largest reported statistical significance for the run I diboson excesses is for the $WZ$ channel, and partly that this involves a more constrained and therefore more interesting symmetry structure than does a simple neutral resonance (though of course what is more interesting can be a matter of perspective).

Here we might point out that there is also the possibility of leakage between the $WZ$, $WW$ and $ZZ$ channels due to misidentification. One interesting work in this regard is \cite{Allanach:2015hba} which carries out a goodness of fit comparison for the various channels (see table V). The 3 fits they compare involve setting one of $WW$ or $WZ$ signal to be zero and fitting the data in terms of the remaining two modes (rows 1 and 2), or by setting the $WW$ and $ZZ$ to be nearly zero and explaining the data almost entirely in terms of $WZ$ (row 3). They find that all 3 fits have $\Delta \chi^2$ values less than 1, though setting $WZ$ to be zero gives a marginally better fit than the one in which $WW$ and $ZZ$ are both set to zero. With the 3 fits being compareable in quality,
the diboson signal could be explained more or less equally well by either of the 3 combinations (i.e. $WZ$ with $ZZ$, $WW$ with $ZZ$ or almost entirely in terms of $WZ$) unless more data allows better discrimination. This means that there is considerable room for misidentification between the various channels, with a $W$ mistaken for a $Z$ and vice versa. With that being so, and with the reported statistical significance of the individual $WZ$ channel being the highest, the diboson excess could be explained entirely by a charged resonance decaying to $WZ$, which is the scenario we focus on through most of this work, though we also briefly address the possibility of an accompanying neutral resonance accounting for the reported $WW$ and $ZZ$ events.

Our strategy will be to follow an effective theory approach. We will consider hypercharged fields that are singlets under the standard model $SU(2)_l$ group with different spin structures (scalar, spin 1 and spin 2) for the possible $2~{\rm TeV}$ particle and construct Lagrangian terms allowed by the symmetries. Since we are assuming $SU(2)$ singlets, the only way for these new fields to get an electric charge is for them to have hypercharge $\pm1$. For each spin case, we will start by assuming that there is no physics in addition to the standard model up to the $2~{\rm TeV}$ range except the possible resonance particle and relax this assumption only if we are forced to do so by some consistency requirements or existing experimental constraints. We will run into such an issue for the vector case where the electroweak precision bounds will force us to include a neutral $Z'$ in addition to $W'$. A $Z'$ could also potentially account for some of the $WW$ and $ZZ$ excess found in the LHC run I data, a possibility we will briefly discuss in the course of our analysis.

It is worth mentioning that in \cite{Kim:2015vba}, a somewhat similar effective theory framework has been used to investigate various spin structures for possible singlet resonances to account for the recently reported diboson anomaly. However, their study is strictly restricted to neutral candidates with the view that the reported $WZ$ excess could well be a $WW$ or $ZZ$ channel being mistaken as $WZ$ due to possible contamination \cite{Allanach:2015hba}, whereas in this paper, we mainly focus on charged resonances. Moreover, in the analysis of a vector resonance \cite{Kim:2015vba} does not take into account electroweak precision bounds which require the introduction of a $W'$ in addition to $Z'$ in order to avoid large deviations of the $\rho$ parameter from unity. Another related work is \cite{Fichet:2015yia} which sets up the effective theory for spin 0 and 2 SM singlet resonances in the context of the diboson anomaly. Yet another alternative is to consider an $SU(2)_l$ triplet with vanishing hypercharge \cite{Thamm:2015csa}. As for the hypercharge case, we would like to acknowledge that \cite{Grojean:2011vu} is one of the earliest papers discussing the phenomenology of a $W'$ using an effective theory approach and even predicted the $WZ$ diboson decay channel back in 2011. We may also mention that some works have also considered explanations other than the diboson interpretation involving a $WW$, $ZZ$ or $WZ$ pair. These include the triboson scenario \cite{Aguilar-Saavedra:2015rna, Aguilar-Saavedra:2015iew, Bhattacherjee:2015svr} or the possibility that some BSM boson with a mass sufficiently close to $m_w$ and $m_z$ may have been misidentified as a $W$ or $Z$ \cite{Chen:2015xql, Allanach:2015blv}.

The organization of this paper will be as follows. In section 2, we consider a hypercharged scalar as a candidate for the possible $2~{\rm TeV}$ resonance. We show that such a scalar cannot account for the diboson anomaly since the symmetries of the standard model prohibit its decay to $WZ$ and $WH$ at tree-level at least up to dimension 5 operators. We also extend the discussion to the case of the 2 higgs doublet model and show that a hypercharged scalar along with the 2HDM cannot account for the $WZ$ excess either. We may also mention here that the 2HDM by itself cannot account for the diboson signal since the tree-level $WZ$ decay of the heavy charged higgs is well-known to be forbidden by the custodial symmetry \cite{Branco:2011iw, Yagyu:2012qp} 
and there are only a few studies where possibilities involving extensions of the 2HDM have been considered \cite{Chen:2015xql, Omura:2015nwa, Sierra:2015zma}.

In section 3, we discuss the possibility of a hypercharged vector $W'$ that quadratically mixes with $W$ as a possible explanation for the diboson signal. The underlying physics for such a vector particle may be an additional gauge field such as that in the $SU(2)_l \times SU(2)_r$ model \cite{Mohapatra:1974hk, Mohapatra:1974gc, Senjanovic:1975rk}
which has also received considerable interest in the context of the diboson anomaly with \cite{Patra:2015bga, Hisano:2015gna, Cheung:2015nha, Dobrescu:2015qna, Gao:2015irw, Brehmer:2015cia, Dev:2015pga, Das:2015ysz, Aguilar-Saavedra:2015iew, Shu:2015cxm, Shu:2016exh} being some especially interesting works. \cite{Berlin:2016hqw} goes a step further by considering the left-right-symmetric model to simultaneously explain the $2~{\rm TeV}$ diboson excess as well as the $750~{\rm GeV}$ diphoton signal. We can of course also consider more complicated extensions of the SM gauge group such as those considered in \cite{Cao:2015lia, Evans:2015cqq, Aydemir:2015oob}. Alternatively, a hypercharged $W'$ may also arise from a composite theory \cite{Low:2015uha, Carmona:2015xaa}. Working in our model independent effective theory approach, we show that a hypercharged $W'$ vector field can indeed account for the observed excess and calculate the relevant cross-section and decay rates. However, this scenario violates electroweak precision bounds on the $\rho$ parameter unless we also introduce a $Z'$ that quadratically mixes with $Z$. We calculate constraints on the $Z'$ mass and the $ZZ'$ mixing based on electroweak precision data.

In section 4, we discuss the hypercharged spin 2 case and show that like the scalar, it too cannot have diboson decays to $WZ$ and $WH$, though the argument for this is slightly different. We thus conclude that within the assumption that there is no additional physics beyond the standard model up to the scale of the possible resonance ($2~{\rm TeV}$ in this case), only a vector resonance can possibly account for the recently reported $WZ$ and $WH$ anomalies, and therefore studies on this subject should focus their efforts accordingly.

\section{Hypercharged lorentz scalar}

We will consider this for the regular standard model as well as its extended version in which there are two higgs doublets and show that a hypercharged scalar cannot account for the diboson excess.

 \subsection{A hypercharged scalar added to the regular standard model}

We start by considering an $SU(2)_l$ singlet  scalar $\phi$ with hyper charge 1 and try to construct interactions that give its decays into $WZ$ and $WH$. Throughout this paper, we will work in the notation where the higgs doublet $H$ transforms as $(2, -1/2)$ under the standard model $SU(2)_l \times U(1)$ group, and acquires a non-zero vacuum expectation value in its first component from electroweak symmetry breaking. With $H$ having hypercharge $-1/2$, we need $\phi$ coupling to two powers of $H$ to get a hypercharge singlet. Additionally, we throw in a pair of covariant derivatives in order to obtain couplings of $\phi$ $WZ$ and $WH$ (in any case, $\phi H\dot H$ is zero). We thus get the dimension 5 interaction
\be
\sL_{\phi hh} 
= -\frac{c}{\Lambda} \phi H\dot D_\mu D^\mu H \, +h.c
\label{d_phi-interaction}
\ee
where $\Lambda$ is the scale associated with the underlying UV physics. This is the only (dimension 5) coupling of $\phi$ to two powers of $H$ since $\phi (D_\mu H) \dot (D^\mu H)$ is zero due to the anti-symmetry of the $SU(2)$ invariant dot product, and $(D_\mu \phi^*) H\dot D^\mu H$ is related to $\phi^* H\dot D_\mu D^\mu H$ through integration by parts.
Naively, if we expand this in terms of the higgs components, we get $\phi W^\mu Z_\mu$ and $(\partial_\mu \phi) W^\mu (H+V)^2$ interactions, in which $V$ is the higgs vacuum expectation value. We may therefore be led to believe that we should get $WZ$ and $WH$ decays of $\phi$. However, if we use the equations of motion for the higgs doublet to eliminate $D_\mu D^\mu H$, we find that (\ref{d_phi-interaction}) is equal to
\be
\frac{c}{\Lambda} 
\left(
Y_u \phi H \dot \bar U_r Q_l
+Y_d \phi \bar Q_l D_r \dot H
+Y_l \phi \bar L_l e_r \dot H
\right)
\,+\, h.c
\label{yukawa-factor-supressed-phiHQQ}
\ee
where $Q_l$ and $L_l$ are the left-handed quark and lepton $SU(2)$ doublets, $Y_u$, $Y_d$ and $Y_l$ are the Yukawa couplings for up and down type quarks and leptons, respectively, and there is an implicit quark generation index (and a CKM matrix for terms in which $u$ type quarks are coupled to $d$ type quarks when we switch to the mass eigen basis). The $\phi W^\mu Z^\mu$ and $(\partial_\mu \phi) W^\mu (H+V)^2$ terms are all gone and we do not get diboson decays of $\phi$ at least at tree-level.

The absence of these decays can also be seen by working carefully with (\ref{d_phi-interaction}). The $(\partial_\mu \phi) W^\mu (H+V)^2$ term contains a mixing between $\phi$ and $W$. This results in an additional set of contributions to the diboson decay amplitude where $\phi$ first flips to a virtual $W$, which then decays to $WZ$ or $WH$ through the standard model $WWZ$ and $WWH$ couplings. And this additional set of contributions (through the virtual $W$) exactly cancel the contributions from the direct $\phi W^\mu Z_\mu$ and $(\partial_\mu \phi) W^\mu H$ interactions due to the custodial symmetry.

We have thus found that a hypercharged scalar, at least by itself, cannot account for the observed anomaly as it does not have the required diboson decays at tree-level up to operators of dimension 5\footnote{We can consider higher dimensional operators like
$\phi (H^\d D^\mu H) (H \dot D_\mu H)$,
which may give the $\phi \to WZ$ decay at tree-level, but of course the decay rate will be highly suppressed.}.
We have not even addressed the other question of getting $pp \to \phi$ with a large enough cross-section. The issue on this front arises from the fact that we are unable to obtain Yukawa interactions between quark bilinears and $\phi$ except through non-renormalizable higgs couplings of the form $\phi H \dot \bar U P_r Q_l$ and $\phi H\dot Q_l P_r D_r$. The Yukawa interactions of $\phi$ to charged quark bilinears thus obtained are suppressed by $V/\Lambda$, which results in very small cross-sections for $pp \to \phi$ even if we are able to do some model building to get the couplings of the first generation quarks to be close to unity. If we try to write couplings of $\phi$ to a pair of right-handed quark fields, then Lorentz-invariance forces us to have currents, and we can only get couplings like $(D_\mu \phi) \bar U_r \gamma^\mu D_r$, which turns out to be further suppressed due to angular momentum conservation). However, at least in principle, it is possible that we might be able to produce $\phi$ from a $pp$ collision in a large enough number to be detectable in a next generation collider if not the LHC. But the absence of diboson decays of $\phi$ means that a stand-alone hypercharged scalar added to the standard model will have to be ruled out as a candidate for explaining any observed diboson signal even in next generation collider experiments.

\subsection{Extending to the 2 higgs doublet model}

We might be tempted to ask whether the above conclusion (i.e. the absence of $WZ$ and $WH$ decays) also holds for the 2 higgs version of the standard model since there, we can also write interactions in which $\phi$ (or its covariant derivative) couples to a product of the two higgs doublets (or their covariant derivatives) rather than the same doublet. We now show by working with the type II 2HDM that the answer is in the affirmative at least for the $WZ$ channel.

For the type II 2HDM, our hypercharged scalar can have the cubic interactions with a pair of higgs fields
\be
\mu_{\phi HH} \phi H_u^\d \dot H_d
\,+\,h.c
\label{phi-higgs-coupling-2hdm}
 \ee
where $H_u$ and $H_d$ transform as $(2, 1/2)$ and $(2, -1/2)$ respectively under the $SU(2) \times U(1)$ gauge group and have the components
\be
H_u = 
\pmatrix{
H^+_u \cr
H^0_u \cr
}
\ee
and
\be
H_d = 
\pmatrix{
H^0_d \cr
H^-_d \cr}
\ee
We can write the neutral components in terms of their vacuum expectation values and real and imaginary parts as
\bea
H^0_u &=& \frac{1}{\sqrt{2}} \left(V_u + X_u + i Y_u \right) \nn \\
H^0_d &=& \frac{1}{\sqrt{2}} \left(V_d + X_d + i Y_d \right)
\eea
where $X_u$, $X_d$, $Y_u$ and $Y_D$ are all real scalar fields, and the vacuum expectation values $v_u$ and $v_d$ satisfy $\sqrt{v_u^2 +v_d^2} = v = 246~{\rm GeV}$. We also define the angle $\beta$ in terms of the equation $\tan\beta = V_u/V_d$.
 
with the neutral components acquiring non-zero vacuum expectation values, (\ref{phi-higgs-coupling-2hdm}) contains a quadratic mixing between $\phi$ and the charged higgs $H^\pm$
\be
\frac{\mu_{\phi HH}}{V} 
\phi H^-
\,+\,h.c
\label{phi-charged-higgs-mixing}
\ee
where $H^\pm$ is the combination
\be
H^\pm = H^\pm_u \cos\beta + H^\pm_d \sin\beta
\ee
We thus have a quadratic mixing through which $\phi$ inherits all the decays of the charged higgs. It is well-known from the literature on the 2HDM that the charged higgs boson does not have a tree-level decay to $WZ$ due to custodial symmetry (see \cite{Branco:2011iw, Yagyu:2012qp} for a good overview). Moreover, there is also no $\phi G^\pm G^0$ term in (\ref{phi-higgs-coupling-2hdm}), where $G^\pm$ and $G^0$ are the goldstone modes associated with the $W^\pm$ and $Z$ bosons, respectively, and are given by
\be
G^\pm = H^\pm_u \sin\beta -H^\pm_d \cos\beta
\ee
and
\be
G^0 = Y_u \sin\beta -Y_d \cos\beta
\ee
Therefore, we conclude that $\phi$ does not have a $WZ$ decay at least at tree-level.

As for $\phi \to WH$, the situation is slightly more subtle since  the neutral scalar states in general have a different diagonalization from the charged and pseudoscalar states. (\ref{phi-higgs-coupling-2hdm}) gives the coupling
\be
\frac{\mu_{\phi HH}}{\sqrt{2}}
\phi G^- (x_d \sin\beta -x_u \cos\beta)
\ee
and unless the linear combination in parentheses is totally orthogonal to the light neutral higgs mode, we do get a $\phi \to WH$ contribution. That said, since the recently observed diboson excesses includes a larger $WZ$ signal, and since $\phi$ added to the 2HDM does not give any tree-level $WZ$ decay, we conclude that the 2HDM cannot account for the $WZ$ excess.

However, this still leaves one more possibility involving the 2HDM which we now very briefly address. What if the quadratic mixing between $\phi$ and the charged higgs creates a heavy mass eigenstate with mass $2~{\rm TeV}$
and a light eigenstate whose mass is somewhere near $m_w$ and $m_z$. Could the observed excess be accounted for by the decay of the heavier eigenstate
to $Z$ and the lighter mode misinterpreted as the $WZ$ channel? A somewhat similar scenario has been proposed in \cite{Chen:2015xql} for the pseudo scalar higgs where it was suggested that if we add a SM gauge singlet complex scalar to the 2HDM, then it is possible to generate mixings between the pseudo scalar component of the singlet with the massive neutral pseudo scalar higgs. If the lighter pseudo-scalar eigen state arising from this mixing has a mass sufficiently close to the $Z$ mass, then the decay of the charged higgs to a $W$ boson along with this lighter pseudo-scalar could potentially have been mistaken as $WZ$. However, while the scenario of the charged higgs of the 2HDM quadratically mixing with $\phi$ to give a light particle which may have been confused as $Z$ may seem appealing, it is not viable since this will also give an overly large contribution to the decay of the top quark to the lighter eigen state.

\section{The vector case}

We now consider a vector field $W'$ with hypercharge $\pm 1$ \cite{Grojean:2011vu}. Such a field can only couple to right-handed fermion currents
\be
g_r \left(W'_\mu \bar U_r \gamma^\mu D_r
\,+\, W'_\mu \bar \nu_r \gamma^\mu e_r\right)
\,+ h.c
\ee
where we have also introduced right-handed neutrinos. For simplicity, we will assume that these interactions are flavour diagonal and all quark generations have the same coupling to $W'$. 

While our goal in this paper is to work in the effective theory framework, let us make some brief comments to motivate that such a theory is indeed possible. For a vector field to have a charge under an abelian gauge field, it either needs to be a non-abelian gauge field itself or a composite particle. The case of a $W'$ being a non-abelian gauge field can for instance arise from a $SU(2)_l \times SU(2)_r \times U(1)$ model
\cite{Mohapatra:1974hk, Mohapatra:1974gc, Senjanovic:1975rk}
where $W'$ is an $SU(2)_r$ gauge field which acts on right-handed fermion $SU(2)_r$ doublets. The higgs field is an $SU(2)_l\times SU(2)_r$ object with 2 of its components acquiring non-zero vacuum expectation values as discussed by \cite{Hisano:2015gna} in the context of the diboson anomaly. The higgs Yukawa terms which give masses to fermions are of the form $H_{i j} \bar f_{L, i} f_{R, j}$, where $L$/$R$ denote left/right handed and $i$ and $j$ are $SU(2)_l$ and $SU(2)_r$ indices. This requires the introduction of right-handed neutrinos in order to account for lepton masses. However, in a limit where one of the higgses is very heavy and can be integrated out, we get an effective theory in which the higgs is just an $SU(2)_l\times U(1)$ doublet and $W'$ is a hypercharged vector with no other symmetry indices. With $W'$ being an $SU(2)$ gauge field, there also has to be a $Z'$, though it is heavier than $W'$ because of $SU(2)_r \times U(1)$ symmetry breaking which also gives the $W'$ its mass.

In the event of $W'$ being a composite field, we do not need to have an $SU(2)_l\times SU(2)_l$ higgs to account for fermion masses, and therefore we start with the regular standard model higgs doublet even in the full theory. One would also generally expect a $Z'$ in the composite case, though now the $W'$ and $Z'$ masses are not produced by the breaking of a gauge symmetry, and have different underlying dynamics. In short, the effective theory for a composite $W'$ and $Z'$ is somewhat similar to the $SU(2)_r\times SU(2)_l$ gauge theory, except that it does not necessitate having right-handed neutrinos at least from any symmetry requirements. It is of course another matter that the right-handed neutrino should be introduced regardless of that because of the non-zero mass for the neutrinos.
 
Having argued that a hypercharged $W'$ is indeed plausible, let us now proceed to discuss its physics. As pointed out by \cite{Hisano:2015gna}, a $W'$ needs to satisfy 2 sets of constraints:
\begin{enumerate}
\item The electroweak precision bounds which constrain the mixing between $W$ and $W'$. This mixing results in deviations of the $\rho$ parameter from unity, and are tightly bound \cite{Peskin:1991sw, delAguila:2010mx, Baak:2014ora}.
\item There are also the Drell-Yan bounds that the production cross-section times leptonic decay branching ratio for $W'$ ($\sigma(pp \to W') \times Br(W'\to LL)$) should be much smaller than $1 ~{\rm fb}$ \cite{Aad:2014cka, ATLAS:2014wra, Khachatryan:2014fba, Khachatryan:2014tva}.
\end{enumerate}

To satisfy the first of these requirements, we will require that $Z'$ not be much heavier than $W'$. This way, the deviations of the $\rho$ parameter from 1 due to the $WW'$ are somewhat offset by effects due to the $ZZ'$ mixing. We will return to this shortly when we introduce $Z'$. As for the Drell-Yan constraints, these are satisfied if the right handed neutrinos are heavier than $W'$. Given that the lower bounds on right-handed neutrino masses are much larger anyway, the Drell-Yan bounds are already satisfied and we will not need to discuss them any further.

Now, coming to the higgs interactions of $W'$, we now write the dimension 4 term
\be
i c_{\pm} W^{'\mu+} H\dot D_\mu H \,+\,h.c
= \frac{c_{\pm} e}{2\sqrt{2} s_W}
W^{'\mu +} W^- _\mu (H+V)^2 \,+\, h.c
\label{WW'-interaction}
 \ee
where we have expanded the higgs doublet in  unitary gauge
\be
H = 
\pmatrix{
\frac{1}{\sqrt{2}} (h+V) \cr
0 \cr
}
\label{higgs-unitary-gauge}
\ee
with $V = 246 GeV$.

This not only contains a quadratic mixing between $W'$ and $W$, but also has a $W' W H$ interaction. 
The $W'\to WH$ decay therefore has 2 contributions. One from the direct coupling and the other through the $WW'$ mixing which flips a $W'$ to a virtual $W$, which in turn  decays to $WH$ through the standard model $WWZ$ or $WWH$ couplings. However, unlike the hypercharge scalar case, these two contributions do not cancel. As for the $WZ$ decay, there is no direct $W'WZ$ coupling and the only tree-level contribution therefore is through a virtual $W$ produced by the $WW'$ mixing.

With $2~{\rm TeV}$ much larger than the $W$ and $Z$ masses, we can work in the limit where $m_w$, $m_z$ and $V$ are very small. This allows us to use the Goldstone equivalence theorem and we get the $W' \to WZ$ decay rate 
\be
\Gamma(W'\to WZ, WH)
\to 
 \frac{m_{w'} c_\pm ^2}{96\pi}
\label{w-prime-wz-decay}
\ee
which for $m_{w'} = 2~{\rm TeV}$ gives $6.63 \, c_\pm^2 \, GeV$.

The decay width for $W'$ to a pair of quarks in the massless quark limit is
\be
\Gamma(W'\to u_i\bar d_j)
= \frac{g_r^2 m_{w'}}{8\pi}
\label{w-prime-qq-decay}
\ee
If $g_r$ is the same as the $W$ coupling to charged quark currents $e_w$, as is usually assumed for the $SU(2)_l \times SU(2)_r$ model to satisfy anomaly cancellation, then this gives $4.09 \, GeV$ for $m_{w'} = 2~{\rm TeV}$.

The $WZ$, $WH$ and $u_i \bar d_j$ channels are the major decay modes of $W'$.
Beyond these, the only other 2 body decay is the $W\gamma$ process, but it is highly suppressed because the photon does not have a longitudinal mode. Therefore, the leading order total decay width comes to about
\be
\Gamma(W') 
= \frac{3g_r^2 m_{w'}}{8\pi}
+ \frac{2m_{w'} c_\pm ^2}{96\pi}
\label{total-w-prime-decay-width}
\ee

Now, coming to the $pp \to W'$ process, we used CT14 PDFs  
\cite{Dulat:2015mca} for calculating the cross-section. For the $8 TeV$ $pp$ center of mass energy, we obtain the cross-sections
\bea
\Sigma_{8~{\rm TeV}}(pp \to W'^{\pm}) 
&= 1440.1 \, g_r^2 \, ~{\rm fb} \\
\Sigma_{13~{\rm TeV}}(pp \to W'^{\pm}) 
&= 11.26 \times 10^3 \, g_r^2 \, ~{\rm fb}
\eea
which for $g_r = e_w$ give $74.06~{\rm fb}$ and $579~{\rm fb}$, respectively\footnote{These cross-sections include both $W'^+$ and $W'^-$ production since both contribute to the diboson signal.}.

From (\ref{w-prime-wz-decay}), (\ref{total-w-prime-decay-width}) and the assumption that $g_r$ is equal to the $W$ coupling to charged standard model fermions, we can obtain the branching ratios for $WZ/WH$ and the cross-sections for $W'$ production in a collision of 2 protons. Table \ref{interesting_c_values} shows some interesting values of $c_\pm$ along with the corresponding branching ratio times cross-sections.

\begin{table}
\begin{tabular}{cccc} 
$|c_\pm|$ & $Br(W' \to WZ)$ & $\Sigma_{8~{\rm TeV}}(pp \to WZ)$ in $fb$ & $\Sigma_{13~{\rm TeV}}(pp \to WZ)$ in $fb$ \\
\hline
1.00 & 0.260 & 19.2 & 150 \\
0.464 & 0.0945 & 7.0 & 54.7 \\
0.385 & 0.0691 & 5.12 & 40.0 \\
 0.193 & 0.0193 & 1.43 & 11.2 \\
\end{tabular}
\caption{Some interesting values of $c_\pm$ along with corresponding branching ratios and cross-section times branching ratios for the $PP\to W' \to WZ$ channel for a $2~{\rm TeV}$ resonance.}
\label{interesting_c_values}
\end{table}

Some comments about the table of $|c_\pm|$ values are in order.
The $19.2~{\rm fb}$ cross-section times branching ratio value for $|c_\pm| =1$ for $8~{\rm TeV}$ falls within the range allowed by the run I $WZ$ data but is clearly ruled out by the run II results at 95$\%$ confidence level. In any case, as \cite{Hisano:2015gna} points out, CMS run I results also put a $7~{\rm fb}$ bound on the $WH$ cross-section times branching ratio \cite{Khachatryan:2015bma}, which through the Goldstone equivalence theorem also imposes the same bound on the $WZ$ cross-section. The next $|c_\pm|$ value of $0.464$ in the table corresponds to this bound. Next is $|c_\pm| = 0.385$, giving the $40~{\rm fb}$ cross-section times branching ratio value for $13~{\rm TeV}$, which is the upper bound according to run II $WZ$ data \cite{ATLAS-CONF-2015-073, CMS:2015nmz}. The run II data for the $WH$ channel, on the other hand, is less constraining and gives an upper bound of $60~{\rm fb}$ \cite{Atlas2WH}, and therefore, we do not include it in our table of interesting data points. Now, as we will shortly see, through the quadratic mixing between $W'$ and $W$ in (\ref{WW'-interaction}), all the above-mentioned values for $c_\pm$ result in a larger shift in $m_w$ than what is permitted by electroweak precision bounds, requiring the simultaneous introduction of a $Z'$ in the theory. The last line shows the threshold value of $|c_\pm| = 0.193$ for which the $\rho$ parameter lies at the boundary of the region allowed by precision data without the inclusion of a $Z'$. This corresponds to a cross-section times branching ratio of about $11.2~{\rm fb}$ for a $13~{\rm TeV}$ experiment. Since this is small but not totally negligible, this means that there is also a considerable region of parameter space where the $Z'$ is much heavier than the $W'$ and therefore does not appear in our effective theory at the $TeV$ or even $10~{\rm TeV}$ scale.

\begin{figure}[h]
\centering
\includegraphics[width=0.8\textwidth]{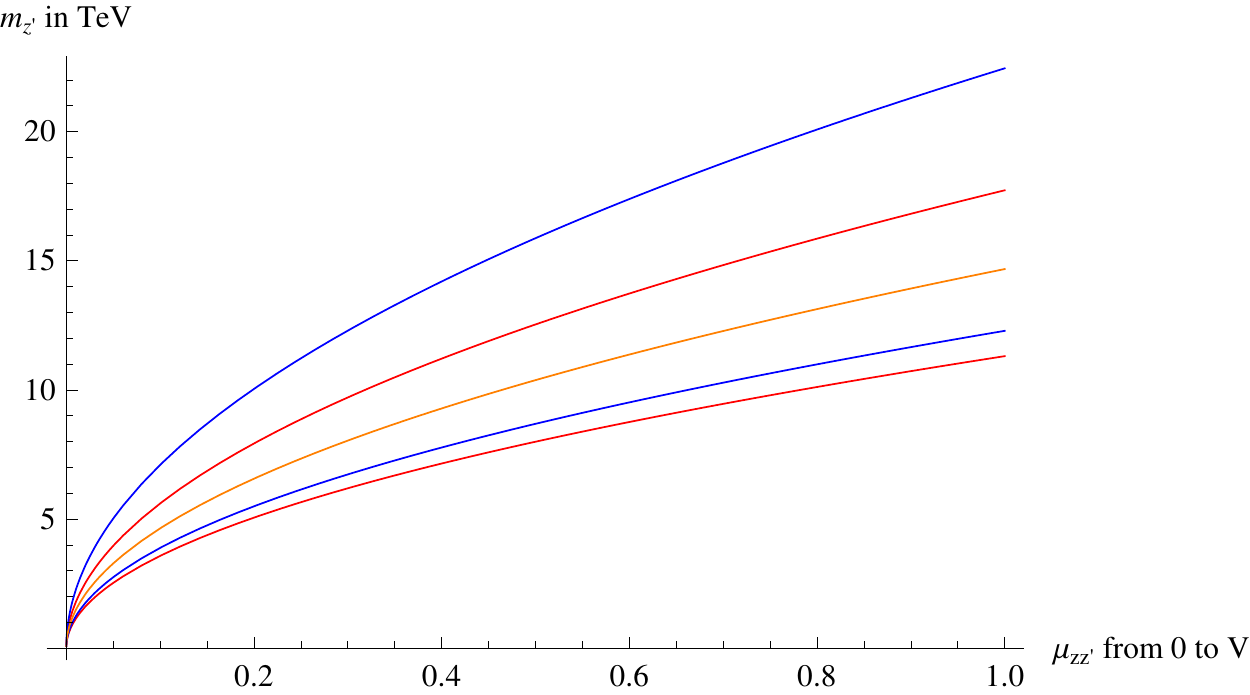}
\caption{Electroweak precision constraints on $m_{z'}$ as a function of $m_{zz'}$ from $0~{\rm to}~V$ for a $2~{\rm TeV}$ $W'$.
The red, blue and orange curves correspond to $WZ$ cross-sections of $54.7~{\rm fb}$, $40~{\rm fb}$ and $11.2~{\rm fb}$, respectively, for a $13~{\rm TeV}$ collision. The red curves correspond to the upper bound for the $WZ$ cross-section from run I data, and the region below these curves is excluded as it pertains to larger cross-sections. The blue curves represent the upper bound on the cross-section set by run II $WZ$ data. The orange curve is the lower bound on $m_{z'}$ for a cross-section of $11.2~{\rm fb}$.
For smaller cross-sections than this, a $z'$ is not needed to satisfy precision constraints.
}
\label{Z-mass-constraint-plot}
\end{figure}

We now address the issue of electroweak precision constraints in some detail and extract bounds on the mass of the $Z'$. The $WW'$ mixing term is
\be
\frac{c_{\pm} e V^2}{2\sqrt{2} s_W}
W^{'\mu +} W^- _\mu \,+\,h.c
= m_w^2 \frac{c_{\pm} s_W \sqrt{2}}{e}
W^{'\mu +} W^- _\mu \,+\,h.c
\ee
where we have taken $m_w^2$ as the tree-level value for the $W$ mass squared, which is equal to $\frac{e^2 V^2}{4 s_W^2}$.
This allows writing the $WW'$ mass matrix as
\be
m_w^2 \pmatrix{
1 & \frac{c_{\pm} s_W \sqrt{2}}{e} \cr
\frac{c_{\pm} s_W \sqrt{2}}{e} & \frac{m_{w'}^2}{m_w^2} \cr
}
\ee
where $m_{w'} = 2~{\rm TeV}$.
By diagonalizing this matrix, we get the leading order percentage shift in the $W$ mass squared
\be
\frac{\Delta m_w^2}{m_w^2}
= -\frac{1}{\frac{m_{w'}^2}{m_w^2} -1}
\frac{2c_{\pm}^2 s_W^2}{e^2}
\label{W-mass-percentage-shift}
\ee
We can now relate this with deviations of the $\rho$ parameter from unity. The $\rho$ parameter is given by
\be
\rho = \frac{m_w ^2}{m_z ^2c_W^2}
\ee
Therefore, in terms of the Peskin-Takeuchi $T$ parameter \cite{Peskin:1991sw}, we get 
\be
\alpha T = \rho -1
= \frac{\Delta m_w^2}{m_w^2} 
-\frac{\Delta m_z^2}{m_z^2} +\ldots 
\label{T-parameter-formula}
\ee
From electroweak precision measurements of the $T$ parameter \cite{Baak:2014ora}, we have $T = 0.10 \pm 0.07$ for $U = 0$. This gives the bounds (since the $95$ percent confidence interval is roughly about $2\sigma$ around the mean), 
\be
-0.04 < T < 0.24
\label{T-bounds}
\ee 
Now, from (\ref{T-parameter-formula}) and (\ref{W-mass-percentage-shift}), we get 
\be
T = -\frac{c^2 s_W^2 m_w^2}{2\alpha^2 \pi (m_{w'}^2 - m_w^2}
\ee
if we assume $\Delta m_z^2 = 0$. And with $m_{w'}^2 = 2~{\rm TeV}$, 
this for any $|c_{\pm}| > 0.193$ is outside the $T$ bounds in (\ref{T-bounds}). Since the more interesting values of $c_\pm$ for explaining the $2~{\rm TeV}$ diboson excess are above this threshold value as shown in table \ref{interesting_c_values}, this means that we must have a $Z'$ lurking nearby with a mixing with $Z$ such that the deviation in $m_z^2$ sufficiently offsets the effect of the shift in the $W$ mass. Specifically, we get the constraint
\be
-0.24 \alpha -\frac{m_w^2 s_W^2 c_{\pm}^2}{2 \alpha \pi (m_{w'}^2-m_w^2)}
< \frac{\Delta m_z^2}{m_z^2}
< 0.04 \alpha -\frac{m_w^2 s_W^2 c_{\pm}^2}{2 \alpha \pi (m_{w'}^2-m_w^2)}
\label{algebraic-constraint-percentage-shift-Z}
\ee
Now, if $Z'$ has a quadratic mixing term with $Z$ of the form
$m_{zz'}^2 Z'_\mu Z^\mu$
, the mass matrix for $Z$ and $Z'$ can be written as
 \be
m_z^2 \pmatrix{
1 & \frac{m_{zz'}^2}{m_z^2} \cr
\frac{m_{zz'}^2}{m_z^2} & \frac{m_{z'}^2}{m_z^2} \cr
}
\ee
and diagonalizing this gives
\be
\frac{\Delta m_z^2}{m_z^2}
= -\frac{1}{\frac{m_{z'}^2}{m_z^2} -1} 
\frac{m_{zz'}^4}{m_z^4}
\label{Z-mass-percentage-shift}
\ee
By combining (\ref{Z-mass-percentage-shift}) with (\ref{algebraic-constraint-percentage-shift-Z}), we obtain bounds on $m_{z'}$ and $m_{zz'}$
which are shown in figure \ref{Z-mass-constraint-plot}.
We focus on $m_{zz'}$ from $0~{\rm to}~V$ to keep the $ZZ'$ mixing small.
The region between the two dashed red curves gives the $m_{z'}$ masses allowed by precision constraints for a given $m_{zz'}$
for $|c_\pm| = 0.464$, corresponding to a $WZ$ cross-section of $7~{\rm fb}$ for a center of mass $PP$ energy of $8~{\rm TeV}$ and about $54.7~{\rm fb}$ for $13~{\rm 
TeV}$. This was the upper bound on the $WZ$ mode from LHC run I. The blue curves on the other hand, give the $m_{z'}$ bounds corresponding to $|c_\pm| = 0.385$, which gives a $WZ$ cross-section times branching ratio of $40~{\rm fb}$ for the $13~{\rm TeV}$ case, which is the upper bound from run II data. The orange curve represents the lower bound on the $Z'$ mass for the threshold value of $c_\pm = 0.193$ below which we do not need to introduce a $Z'$ in the theory in order to satisfy precision constraints. This corresponds to a $WZ$ cross-section times branching ratio of $\sigma_{WZ}$ = $11.2~{\rm fb}$ for a $13~{\rm TeV}$ collision. For any cross-sections smaller than this value, the $z'$ mass must lie somewhere in the region above the orange curve, and this includes the uninteresting scenario that the recently reported excesses do not correspond to any new particle. The region below the red curves is disallowed even by run I. The region below the blue curves is ruled out at 95$\%$ confidence level by the run II data. The combined bound curves therefore lie somewhere in the narrow regions between the red and blue curves\footnote{That is, the lower $m_{z'}$ bound curve corresponding to the combined bound on the cross-times branching ratio will be somewhere between the lower red and blue curves, and the combined upper bound would be somewhere between the upper red and blue curves.}.

We can see that these precision constraints on the $Z'$ mass leave open a wide range of possibilities. For example, a $Z'$ in the $2-4~{\rm TeV}$
range which could potentially be detected at the LHC is very much consistent with the recently reported diboson anomaly. Such a $Z'$ that is slightly heavier than $W'$ could for instance arise from the left-right symmetric model. Interestingly, CMS did report an electron-positron excess at $2.9~{\rm TeV}$ \cite{CMS-DP-2015-039} in run I, though this was a very small event and taking it too seriously may be somewhat premature at this stage. There is also a large part of open parameter space where $Z'$ can be considerably heavier and therefore difficult to detect at the LHC, as well as the somewhat less likely region from the point of view of model building in which it may be lighter than $2~{\rm TeV}$.

Then there is the possibility of a $2~{\rm TeV}$, which could also account for some part of the diboson excess with the somewhat bizarre miracle of $W'$ and $Z'$ masses being the same \footnote{The existance of a neutral resonance with the same mass would not be such a miracle if we were considering an $SU(2)_l$ triplet but in this paper we are restricting our attention to $SU(2)_l$ singlets with hypercharge.}. In this case, all the three modes, namely $WZ$, $WW$, and $ZZ$ would be present in the actual physics. However, as we mentioned in the introduction, there is also considerable room for misidentification between the various channels due to the closeness of the $W$ and $Z$ masses, and the analysis of \cite{Allanach:2015hba}
shows that fitting the data entirely in terms of $WZ$ also provides a reasonably good fit with $\Delta\chi^2$ of $0.8$. For this reason, we do not necessarily need a $2~{\rm TeV}$ to explain the diboson excess. However, taking one of the $WZ$ or $ZZ$ signals to be zero also provides fits of nearly similar quality, and therefore, it is also possible that the diboson signal could be coming entirely from a neutral $Z'$ \cite{Kim:2015vba} or through a mixture of mass degenerate $W'$ and $Z'$ particles decay into all the various diboson channels. That said, having a $W'$ and a $Z'$ with the same mass may require some model building as it is not entirely clear how such a scenario may arise.

\begin{figure}[h]
\centering
\includegraphics[width=0.8\textwidth]{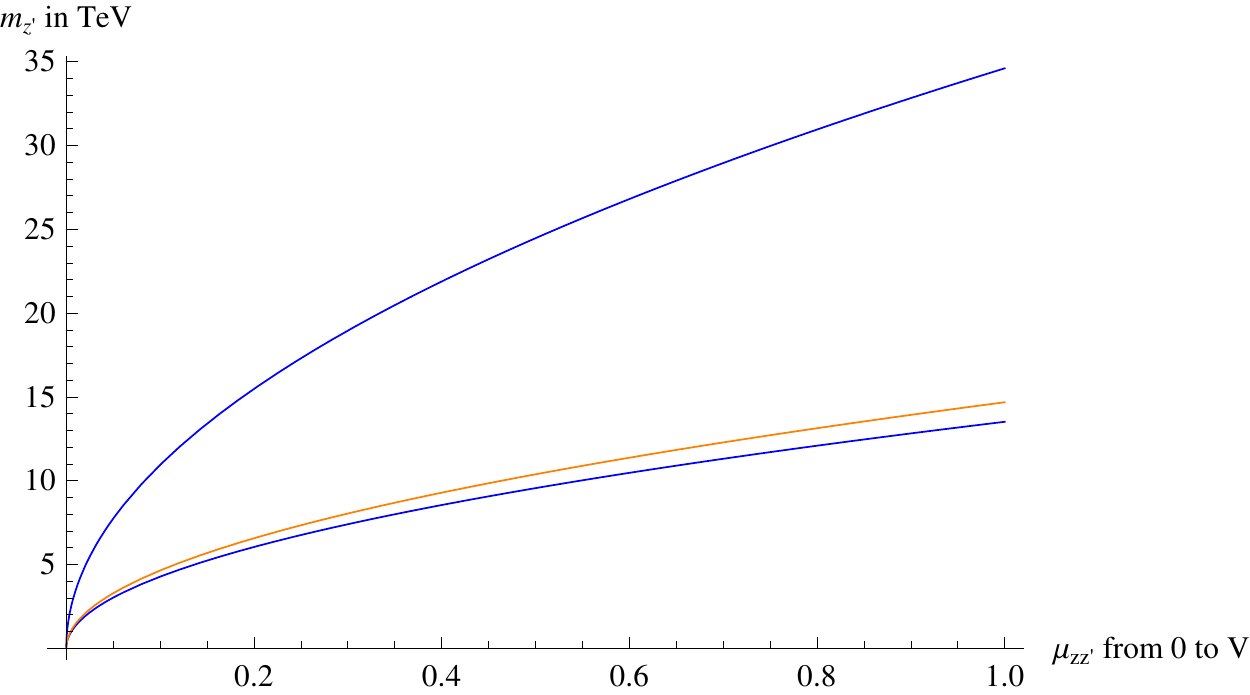}
\caption{Electroweak precision constraints on $m_{z'}$  as a function of $m_{zz'}$ from $0~{\rm to}~V$ for a $1.6~{\rm TeV}$ $W'$.
The blue curves represent the lower and upper bounds on $m_{z'}$ for a $\sigma_{WZ}$ of $50~{\rm fb}$, which is the upper bound set by run II $WH$ data, and the region below these curves is disallowed as it represents larger cross-sections. The orange curve is the lower bound on $m_{z'}$ for a cross-section of $22.8~{\rm fb}$. For any cross-sections smaller than this value, a $z'$ is not needed to satisfy precision constraints, and if a $z'$ does exist, then $m_{z'}$ must lie above the orange curve.
}
\label{Z-mass-constraint-plot-1600-GeV}
\end{figure}

\begin{figure}[h]
\centering
\includegraphics[width=0.8\textwidth]{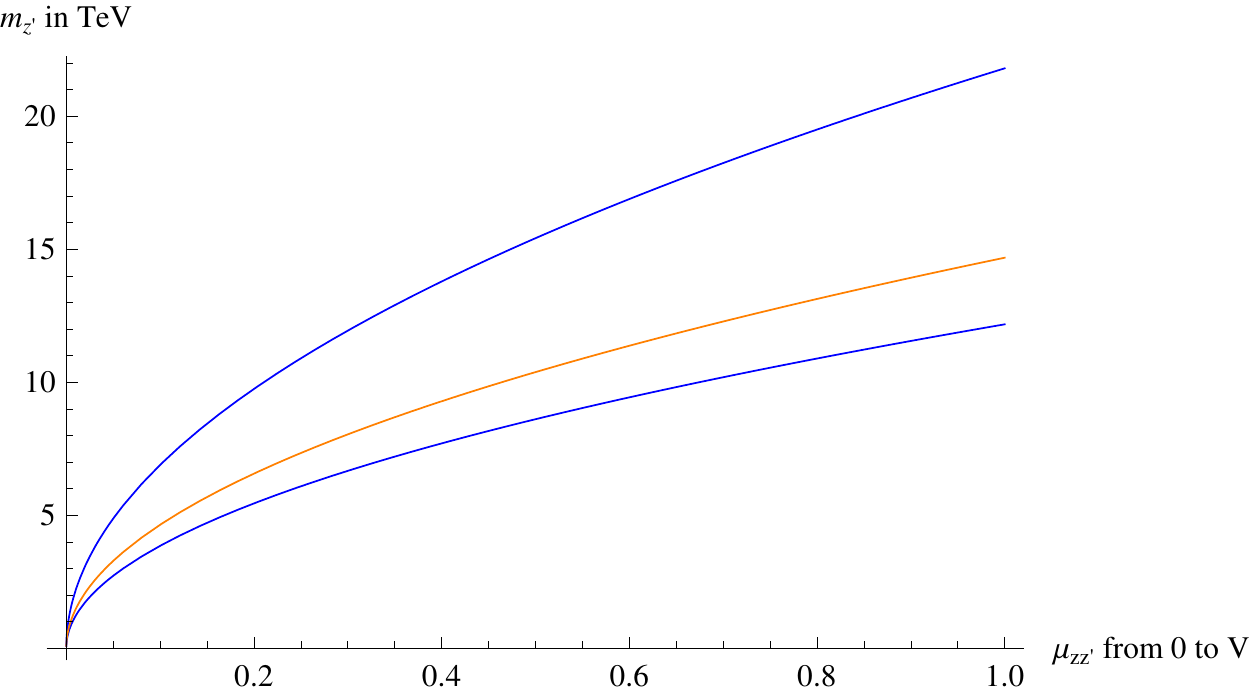}
\caption{Electroweak precision constraints on $m_{z'}$  as a function of $m_{zz'}$ from $0~{\rm to}~V$ for a $2.4~{\rm TeV}$ $W'$.
The blue curves represent the lower and upper bounds on $m_{z'}$ for a $\sigma_{WZ}$ of $20~{\rm fb}$, which is the upper bound set by run II $WH$ data, and the region below these curves is disallowed as it represents larger cross-sections. The orange curve is the lower bound on $m_{z'}$ for a cross-section of $5.62~{\rm fb}$. For any cross-sections smaller than this value, a $z'$ is not needed to satisfy precision constraints, and if a $z'$ does exist, then $m_{z'}$ must lie above the orange curve.
}
\label{Z-mass-constraint-plot-2400-GeV}
\end{figure}

While the primary focus of this paper is the $2~{\rm TeV}$ excess found in LHC run I, our analysis is of course also applicable to any other value of the resonance. We therefore also show precision bounds on $m_{z'}$ for $m_{w'} = 1.6~{\rm TeV}$ and $2.4~{\rm TeV}$ In figures \ref{Z-mass-constraint-plot-1600-GeV} and \ref{Z-mass-constraint-plot-2400-GeV}, respectively, just to illustrate how this works for 2 other $W'$ masses. While neither run of the LHC has found a noticeable excess at these values thus far, the tightest constraints come from run II $WH$ channel data, which gives upper bounds of $50~{\rm fb}$ and $20~{\rm fb}$, respectively, for the cross-section times branching ratios for these two masses for a $W'$ particle \cite{Atlas2WH}. In each of these plots, we show bounds on $m_{z'}$ with a blue pair of curves for $\sigma_{WZ}$ corresponding to the above-mentioned upper bounds set by the run 2 $WH$ data, and the orange line represents the threshold value of $|c_{\pm}|$ below which we do not need to introduce a $Z'$ in the theory in order to satisfy precision constraints.
These threshold values of $|c_\pm|$ correspond to cross-section times branching ratios of $22.8~{\rm fb}$
and $5.62-20~{\rm fb}$ for $1.6~{\rm TeV}$ and $2.4~{\rm TeV}$, respectively, for a $13~{\rm TeV}$ experiment.
We can see that even though no noticeable excess has been reported for these values of the $w'$ resonance, there is still a considerable region of parameter space that remains open.

We conclude this section by listing down the dimension 4 interactions of $Z'$ allowed by symmetries. Continuing with our effective theory approach, we take $Z'$ to be a standard model gauge singlet to make it have no electromagnetic charge. We find that the couplings of $Z'$ to standard model fermions 
are somewhat less constrained than those of $W'$ as $Z'$ can couple to both left and right-handed fermions \cite{Kim:2015vba}
\be
\bar f_i \gamma^\mu Z'_\mu (c_l P_l + c_r P_r) f_i
\ee
 where $i$ is an index labling the various fermions in the standard model. As for the quadratic mixing of $Z'$ with $Z$, the symmetries allow 2 different mechanisms. One of these is kinetic mixing with the hypercharge gauge field as also noted by \cite{Kim:2015vba} 
\be
-\frac{1}{4} B_{\mu\nu} B^{\mu\nu}
-\frac{1}{4} Z'_{\mu\nu} Z'^{\mu\nu}
-\frac{\kappa}{2} Z'_{\mu\nu} B^{\mu\nu}
\label{kinetic-mixing}
\ee 
However, there is also the $Z'$ coupling to the higgs current which has not been considered in \cite{Kim:2015vba} 
\be
i c_0 Z' H^\d D_\mu H
= -\frac{c_0 e}{4 c_W s_W}
Z'_\mu Z^\mu (V+H)^2
\label{Z'-higgs-current-interaction}
\ee
and directly gives a mass mixing of the form $m_{zz'}^2 Z'_\mu Z^\mu$ when we replace the higgs fields with their vacuum expectation values.
The former changes the kinetic energy and the latter directly modifies the mass matrix for the $Z$ and $Z'$. Since simultaneously diagonalizing the kinetic energy and mass terms is rather complicated, we can follow a two-step process. First, we can diagonalize (\ref{kinetic-mixing}) and rescale $B_\mu$ by $\sqrt{1-\kappa^2}$ to obtain canonically normalized kinetic energy terms. We can then diagonalize the mass term in the next step. Since a detailed analysis of the parameter space is beyond the scope of this paper, we will not carry out this procedure here.

We end our discussion of the vector case by noting that the above two quadratic mixing terms with $Z$ results in various diboson decay channels such as $WW$, $ZZ$, $HH$ and $ZH$ as we have mentioned earlier. This not only means that a $Z'$ could possibly also explain the $WW$ and $ZZ$ events in the $2~{\rm TeV}$ diboson excess, but also that searches for neutral diboson resonances should therefore be an integral part of any program for understanding the recently reported diboson anomalies.

\section{The spin 2 case}

The Lagrangian for a massive spin 2 field is the same as the massive graviton (see \cite{Hinterbichler:2011tt} for an excellent review). The standard practise for gravity is to expand the metric around the Minkowski metric or some other static background as $g_{\mu\nu} = \eta_{\mu\nu} + h_{\mu\nu}$. The dynamics of the graviton are then described by $h_{\mu\nu}$. In this paper, we will denote our hypercharged spin 2 field by $\Pi_{\mu\nu}$ in place of $h_{\mu\nu}$ to avoid confusion with the higgs. Now, if we follow our recipe of coupling our hypercharged fields with two powers of the higgs, we find that we are not able to write down any non-zero interactions. Since $\Pi^{\mu\nu}$ is symmetric, $\Pi^{\mu\nu} (D_\mu H) \dot D_\nu H$ is zero due to the anti-symmetry of the $SU(2)$ invariant dot product. The other possible terms to consider are $(D_\mu \Pi^{\mu\nu}) H\dot D_\nu H$
and $(D_\mu \Pi^{\mu\nu}) (D_\nu H)\dot H$, which are in fact related through integration by parts. Now, it is well-known in the literature on massive gravity (see the appendix for a quick derivation) that
\be
D_\mu \Pi^{\mu\nu} = 0
\ee
We are therefore forced to conclude that the diboson anomaly cannot be explained by a hypercharged spin 2 resonance.

\section{Conclusion}

We have carried out a detailed effective theory analysis of hypercharged fields with various spin structures to investigate what type of particles could potentially account for the recently reported diboson excess. Working within the assumption that there is no additional physics beyond the standard model up to the scale of the possible diboson resonance, we have shown that a hypercharged scalar and a spin 2 particle do not have $WZ$ and $WH$ decay channels at tree-level (up to operators of at least dimension 5) and must therefore be ruled out as viable explanations for the anomaly. On the other hand, a hypercharged vector that quadratically mixes with $W$ not only has the required diboson decays but can also have a production cross-section in the right range to account for the $WZ$ and $WH$ excesses.

However, electroweak precision bounds require that such a $W'$ be accompanied by a $Z'$ that quadratically mixes with $Z$. We have calculated constraints on the $Z'$ and its quadratic mixing with $Z$. 
These constraints allow the possibility of a $Z'$ that is slightly heavier than $W'$ as predicted by the $SU(2)_r\times SU(2)_l$ model, but also allow for a heavier $Z$ that may be difficult to detect at the LHC. There is also an open region of parameter space in which $Z'$ can be $2.0~{\rm TeV}$ or lighter, though it is not entirely clear if it is possible to come up with a model with such a spectrum.

Like $W'$, $Z'$ too should have diboson decay modes due to its quadratic mixing with $Z$, except that these will involve the pairs $WW$, $ZZ$, $HH$ and $ZH$. The search for diboson signals can therefore serve as a very useful probe of new physics which will be of relevance even beyond the recently reported diboson excesses.

\section*{Acknowledgments}

The author is especially grateful to Matthew Reece for his guidance and support throughout this project. Special thanks also to Prateek Agrawal, Prahar Mitra, Sabrina Pasterski, Abhishek Pathak, Matthew Schwartz and Taizan Watari for very helpful discussions.

\appendix
\section{Derivation of $D_\mu \Pi^{\mu\nu} = 0$ for a spin 2 field}

Here we give a quick derivation of the equation $D_\mu \Pi^{\mu\nu} =0$ for a massive spin 2 field, which is well-known to experts on massive gravity but may not be familiar to readers outside that field. Readers interested in learning more on the subject may refer to \cite{Hinterbichler:2011tt} for a detailed review.

The Lagrangian for a massive spin 2 field is the same as a massless graviton with the addition of the Fierz-Pauli mass term which is given by
\be
\frac{m^2}{2}
\left(
(\eta^{\mu\nu} \Pi_{\mu\nu})^2
-\Pi^{\mu\nu} \Pi_{\mu\nu}
\right)
\ee
The equations of motion for $\Pi^{\mu\nu}$ are
\be
D^2 \Pi_{\mu\nu}
-D_\lambda D_\mu \Pi^\Lambda _\nu
-D_\lambda D_\nu \Pi^\lambda _\mu
+\eta_{\mu\nu} D_\lambda D_\sigma \Pi^{\lambda\sigma}
+D_\mu D_\nu \Pi
-\eta_{\mu\nu} D^2 \Pi
-m^2(\Pi_{\mu\nu} -\eta_{\mu\nu} \Pi)
= 0
\ee  
where $\Pi$ is the trace $\Pi^\mu _\mu$ and $D^2 = D_\mu D^\mu$.
Acting on this with $D^\mu$, we get for non-zero $m^2$
\be
m^2 (D_\mu \Pi^{\mu\nu} -D_\nu \Pi)
= 0
\label{derivative-of-equation-of-motion}
\ee  
Inserting this back into the equation of motion gives
\be
D^2 \Pi_{\mu\nu}
-D_\mu D_\nu \Pi
-m^2(\Pi_{\mu\nu} -\eta_{\mu\nu} \Pi)
= 0
\ee
Taking the trace of this gives $\Pi = 0$.
And plugging this result in (\ref{derivative-of-equation-of-motion}) gives
\be
D_\mu \Pi^{\mu\nu} = 0
\ee

{\footnotesize
\bibliography{ref}

\providecommand{\href}[2]{#2}\begingroup\raggedright\begin{thebibliography}{10}

\bibitem{Aad:2015owa}
{\bfseries ATLAS} Collaboration, G.~Aad {\em et~al.}, ``{Search for high-mass
  diboson resonances with boson-tagged jets in proton-proton collisions at
  $\sqrt{s}$ = 8 TeV with the ATLAS detector},''
\href{http://arxiv.org/abs/1506.00962}{{\ttfamily arXiv:1506.00962 [hep-ex]}}.

\bibitem{Khachatryan:2014hpa}
{\bfseries CMS} Collaboration, V.~Khachatryan {\em et~al.}, ``{Search for
  massive resonances in dijet systems containing jets tagged as W or Z boson
  decays in pp collisions at $ \sqrt{s} $ = 8 TeV},''
  \href{http://dx.doi.org/10.1007/JHEP08(2014)173}{{\em JHEP} {\bfseries 08}
  (2014) 173},
\href{http://arxiv.org/abs/1405.1994}{{\ttfamily arXiv:1405.1994 [hep-ex]}}.

\bibitem{CMS:2015gla}
{\bfseries CMS} Collaboration, C.~Collaboration,
``{Search for massive WH resonances decaying to $\ell \nu {\rm b \bar{b}}$
  final state in the boosted regime at $\sqrt{s}=8$\,TeV},''.

\bibitem{ATLAS-CONF-2015-073}
``{Search for resonances with boson-tagged jets in 3.2 fb?1 of p p collisions
  at ? s = 13 TeV collected with the ATLAS detector},'' Tech. Rep.
  ATLAS-CONF-2015-073, CERN, Geneva, Dec, 2015.
\newblock \url{http://cds.cern.ch/record/2114845}.

\bibitem{CMS:2015nmz}
{\bfseries CMS} Collaboration, C.~Collaboration,
``{Search for massive resonances decaying into pairs of boosted W and Z bosons
  at $\sqrt{s}$ = 13 TeV},''.

\bibitem{Hisano:2015gna}
J.~Hisano, N.~Nagata, and Y.~Omura, ``{Interpretations of the ATLAS Diboson
  Resonances},'' \href{http://dx.doi.org/10.1103/PhysRevD.92.055001}{{\em Phys.
  Rev.} {\bfseries D92} no.~5, (2015) 055001},
\href{http://arxiv.org/abs/1506.03931}{{\ttfamily arXiv:1506.03931 [hep-ph]}}.

\bibitem{Khachatryan:2015bma}
{\bfseries CMS} Collaboration, V.~Khachatryan {\em et~al.}, ``{Search for A
  Massive Resonance Decaying into a Higgs Boson and a W or Z Boson in Hadronic
  Final States in Proton-Proton Collisions at $\sqrt{s}$ = 8 TeV},''
\href{http://arxiv.org/abs/1506.01443}{{\ttfamily arXiv:1506.01443 [hep-ex]}}.

\bibitem{Allanach:2015hba}
B.~C. Allanach, B.~Gripaios, and D.~Sutherland, ``{Anatomy of the ATLAS diboson
  anomaly},'' \href{http://dx.doi.org/10.1103/PhysRevD.92.055003}{{\em Phys.
  Rev.} {\bfseries D92} no.~5, (2015) 055003},
\href{http://arxiv.org/abs/1507.01638}{{\ttfamily arXiv:1507.01638 [hep-ph]}}.

\bibitem{Kim:2015vba}
D.~Kim, K.~Kong, H.~M. Lee, and S.~C. Park, ``{ATLAS Diboson Excesses
  Demystified in Effective Field Theory Approach},''
\href{http://arxiv.org/abs/1507.06312}{{\ttfamily arXiv:1507.06312 [hep-ph]}}.

\bibitem{Fichet:2015yia}
S.~Fichet and G.~von Gersdorff, ``{Effective theory for neutral resonances and
  a statistical dissection of the ATLAS diboson excess},''
\href{http://arxiv.org/abs/1508.04814}{{\ttfamily arXiv:1508.04814 [hep-ph]}}.

\bibitem{Thamm:2015csa}
A.~Thamm, R.~Torre, and A.~Wulzer, ``{Composite Heavy Vector Triplet in the
  ATLAS Diboson Excess},''
  \href{http://dx.doi.org/10.1103/PhysRevLett.115.221802}{{\em Phys. Rev.
  Lett.} {\bfseries 115} no.~22, (2015) 221802},
\href{http://arxiv.org/abs/1506.08688}{{\ttfamily arXiv:1506.08688 [hep-ph]}}.

\bibitem{Grojean:2011vu}
C.~Grojean, E.~Salvioni, and R.~Torre, ``{A weakly constrained W' at the early
  LHC},'' \href{http://dx.doi.org/10.1007/JHEP07(2011)002}{{\em JHEP}
  {\bfseries 07} (2011) 002},
\href{http://arxiv.org/abs/1103.2761}{{\ttfamily arXiv:1103.2761 [hep-ph]}}.

\bibitem{Aguilar-Saavedra:2015rna}
J.~A. Aguilar-Saavedra, ``{Triboson interpretations of the ATLAS diboson
  excess},'' \href{http://dx.doi.org/10.1007/JHEP10(2015)099}{{\em JHEP}
  {\bfseries 10} (2015) 099},
\href{http://arxiv.org/abs/1506.06739}{{\ttfamily arXiv:1506.06739 [hep-ph]}}.

\bibitem{Aguilar-Saavedra:2015iew}
J.~A. Aguilar-Saavedra and F.~R. Joaquim, ``{Multiboson production in W'
  decays},''
\href{http://arxiv.org/abs/1512.00396}{{\ttfamily arXiv:1512.00396 [hep-ph]}}.

\bibitem{Bhattacherjee:2015svr}
B.~Bhattacherjee, P.~Byakti, C.~K. Khosa, J.~Lahiri, and G.~Mendiratta,
  ``{Alternative search strategies for a BSM resonance fitting ATLAS diboson
  excess},''
\href{http://arxiv.org/abs/1511.02797}{{\ttfamily arXiv:1511.02797 [hep-ph]}}.

\bibitem{Chen:2015xql}
C.-H. Chen and T.~Nomura, ``{2 TeV Higgs boson and diboson excess at the
  LHC},'' \href{http://dx.doi.org/10.1016/j.physletb.2015.08.028}{{\em Phys.
  Lett.} {\bfseries B749} (2015) 464--468},
\href{http://arxiv.org/abs/1507.04431}{{\ttfamily arXiv:1507.04431 [hep-ph]}}.

\bibitem{Allanach:2015blv}
B.~C. Allanach, P.~S.~B. Dev, and K.~Sakurai, ``{The ATLAS Di-boson Excess
  Could Be an $R-$parity Violating Di-smuon Excess},''
\href{http://arxiv.org/abs/1511.01483}{{\ttfamily arXiv:1511.01483 [hep-ph]}}.

\bibitem{Branco:2011iw}
G.~C. Branco, P.~M. Ferreira, L.~Lavoura, M.~N. Rebelo, M.~Sher, and J.~P.
  Silva, ``{Theory and phenomenology of two-Higgs-doublet models},''
  \href{http://dx.doi.org/10.1016/j.physrep.2012.02.002}{{\em Phys. Rept.}
  {\bfseries 516} (2012) 1--102},
\href{http://arxiv.org/abs/1106.0034}{{\ttfamily arXiv:1106.0034 [hep-ph]}}.

\bibitem{Yagyu:2012qp}
K.~Yagyu, {\em {Studies on Extended Higgs Sectors as a Probe of New Physics
  Beyond the Standard Model}}.
\newblock PhD thesis, Toyama U., 2012.
\newblock \href{http://arxiv.org/abs/1204.0424}{{\ttfamily arXiv:1204.0424
  [hep-ph]}}.
\newblock
\url{http://inspirehep.net/record/1097017/files/arXiv:1204.0424.pdf}.
\newblock

\bibitem{Omura:2015nwa}
Y.~Omura, K.~Tobe, and K.~Tsumura, ``{Survey of Higgs interpretations of the
  diboson excesses},'' \href{http://dx.doi.org/10.1103/PhysRevD.92.055015}{{\em
  Phys. Rev.} {\bfseries D92} no.~5, (2015) 055015},
\href{http://arxiv.org/abs/1507.05028}{{\ttfamily arXiv:1507.05028 [hep-ph]}}.

\bibitem{Sierra:2015zma}
D.~Aristizabal~Sierra, J.~Herrero-Garcia, D.~Restrepo, and A.~Vicente,
  ``{Diboson anomaly: heavy Higgs resonance and QCD vector-like exotics},''
\href{http://arxiv.org/abs/1510.03437}{{\ttfamily arXiv:1510.03437 [hep-ph]}}.

\bibitem{Mohapatra:1974hk}
R.~N. Mohapatra and J.~C. Pati, ``{Left-Right Gauge Symmetry and an
  Isoconjugate Model of CP Violation},''
\href{http://dx.doi.org/10.1103/PhysRevD.11.566}{{\em Phys. Rev.} {\bfseries
  D11} (1975) 566--571}.

\bibitem{Mohapatra:1974gc}
R.~N. Mohapatra and J.~C. Pati, ``{A Natural Left-Right Symmetry},''
\href{http://dx.doi.org/10.1103/PhysRevD.11.2558}{{\em Phys. Rev.} {\bfseries
  D11} (1975) 2558}.

\bibitem{Senjanovic:1975rk}
G.~Senjanovic and R.~N. Mohapatra, ``{Exact Left-Right Symmetry and Spontaneous
  Violation of Parity},''
\href{http://dx.doi.org/10.1103/PhysRevD.12.1502}{{\em Phys. Rev.} {\bfseries
  D12} (1975) 1502}.

\bibitem{Patra:2015bga}
S.~Patra, F.~S. Queiroz, and W.~Rodejohann, ``{Stringent Dilepton Bounds on
  Left-Right Models using LHC data},''
  \href{http://dx.doi.org/10.1016/j.physletb.2015.11.009}{{\em Phys. Lett.}
  {\bfseries B752} (2016) 186--190},
\href{http://arxiv.org/abs/1506.03456}{{\ttfamily arXiv:1506.03456 [hep-ph]}}.

\bibitem{Cheung:2015nha}
K.~Cheung, W.-Y. Keung, P.-Y. Tseng, and T.-C. Yuan, ``{Interpretations of the
  ATLAS Diboson Anomaly},''
\href{http://arxiv.org/abs/1506.06064}{{\ttfamily arXiv:1506.06064 [hep-ph]}}.

\bibitem{Dobrescu:2015qna}
B.~A. Dobrescu and Z.~Liu, ``{A W' boson near 2 TeV: predictions for Run 2 of
  the LHC},''
\href{http://arxiv.org/abs/1506.06736}{{\ttfamily arXiv:1506.06736 [hep-ph]}}.

\bibitem{Gao:2015irw}
Y.~Gao, T.~Ghosh, K.~Sinha, and J.-H. Yu, ``{SU(2)×SU(2)×U(1) interpretations
  of the diboson and Wh excesses},''
  \href{http://dx.doi.org/10.1103/PhysRevD.92.055030}{{\em Phys. Rev.}
  {\bfseries D92} no.~5, (2015) 055030},
\href{http://arxiv.org/abs/1506.07511}{{\ttfamily arXiv:1506.07511 [hep-ph]}}.

\bibitem{Brehmer:2015cia}
J.~Brehmer, J.~Hewett, J.~Kopp, T.~Rizzo, and J.~Tattersall, ``{Symmetry
  Restored in Dibosons at the LHC?},''
  \href{http://dx.doi.org/10.1007/JHEP10(2015)182}{{\em JHEP} {\bfseries 10}
  (2015) 182},
\href{http://arxiv.org/abs/1507.00013}{{\ttfamily arXiv:1507.00013 [hep-ph]}}.

\bibitem{Dev:2015pga}
P.~S. Bhupal~Dev and R.~N. Mohapatra, ``{Unified explanation of the $eejj$,
  diboson and dijet resonances at the LHC},''
  \href{http://dx.doi.org/10.1103/PhysRevLett.115.181803}{{\em Phys. Rev.
  Lett.} {\bfseries 115} no.~18, (2015) 181803},
\href{http://arxiv.org/abs/1508.02277}{{\ttfamily arXiv:1508.02277 [hep-ph]}}.

\bibitem{Das:2015ysz}
K.~Das, T.~Li, S.~Nandi, and S.~K. Rai, ``{The Diboson Excesses in an Anomaly
  Free Leptophobic Left-Right Model},''
\href{http://arxiv.org/abs/1512.00190}{{\ttfamily arXiv:1512.00190 [hep-ph]}}.

\bibitem{Shu:2015cxm}
J.~Shu and J.~Yepes, ``{Left-right non-linear dynamical Higgs},''
\href{http://arxiv.org/abs/1512.09310}{{\ttfamily arXiv:1512.09310 [hep-ph]}}.

\bibitem{Shu:2016exh}
J.~Shu and J.~Yepes, ``{Diboson excess and $Z'$-predictions via left-right
  non-linear Higgs},''
\href{http://arxiv.org/abs/1601.06891}{{\ttfamily arXiv:1601.06891 [hep-ph]}}.

\bibitem{Berlin:2016hqw}
A.~Berlin, ``{The Diphoton and Diboson Excesses in a Left-Right Symmetric
  Theory of Dark Matter},''
\href{http://arxiv.org/abs/1601.01381}{{\ttfamily arXiv:1601.01381 [hep-ph]}}.

\bibitem{Cao:2015lia}
Q.-H. Cao, B.~Yan, and D.-M. Zhang, ``{Simple non-Abelian extensions of the
  standard model gauge group and the diboson excesses at the LHC},''
  \href{http://dx.doi.org/10.1103/PhysRevD.92.095025}{{\em Phys. Rev.}
  {\bfseries D92} no.~9, (2015) 095025},
\href{http://arxiv.org/abs/1507.00268}{{\ttfamily arXiv:1507.00268 [hep-ph]}}.

\bibitem{Evans:2015cqq}
J.~L. Evans, N.~Nagata, K.~A. Olive, and J.~Zheng, ``{The ATLAS Diboson
  Resonance in Non-Supersymmetric SO(10)},''
\href{http://arxiv.org/abs/1512.02184}{{\ttfamily arXiv:1512.02184 [hep-ph]}}.

\bibitem{Aydemir:2015oob}
U.~Aydemir, ``{SO(10) grand unification in light of recent LHC searches and
  colored scalars at the TeV-scale},''
\href{http://arxiv.org/abs/1512.00568}{{\ttfamily arXiv:1512.00568 [hep-ph]}}.

\bibitem{Low:2015uha}
M.~Low, A.~Tesi, and L.-T. Wang, ``{Composite spin-1 resonances at the LHC},''
  \href{http://dx.doi.org/10.1103/PhysRevD.92.085019}{{\em Phys. Rev.}
  {\bfseries D92} no.~8, (2015) 085019},
\href{http://arxiv.org/abs/1507.07557}{{\ttfamily arXiv:1507.07557 [hep-ph]}}.

\bibitem{Carmona:2015xaa}
A.~Carmona, A.~Delgado, M.~Quirós, and J.~Santiago, ``{Diboson resonant
  production in non-custodial composite Higgs models},''
  \href{http://dx.doi.org/10.1007/JHEP09(2015)186}{{\em JHEP} {\bfseries 09}
  (2015) 186},
\href{http://arxiv.org/abs/1507.01914}{{\ttfamily arXiv:1507.01914 [hep-ph]}}.

\bibitem{Peskin:1991sw}
M.~E. Peskin and T.~Takeuchi, ``{Estimation of oblique electroweak
  corrections},''
\href{http://dx.doi.org/10.1103/PhysRevD.46.381}{{\em Phys. Rev.} {\bfseries
  D46} (1992) 381--409}.

\bibitem{delAguila:2010mx}
F.~del Aguila, J.~de~Blas, and M.~Perez-Victoria, ``{Electroweak Limits on
  General New Vector Bosons},''
  \href{http://dx.doi.org/10.1007/JHEP09(2010)033}{{\em JHEP} {\bfseries 09}
  (2010) 033},
\href{http://arxiv.org/abs/1005.3998}{{\ttfamily arXiv:1005.3998 [hep-ph]}}.

\bibitem{Baak:2014ora}
{\bfseries Gfitter Group} Collaboration, M.~Baak, J.~Cúth, J.~Haller,
  A.~Hoecker, R.~Kogler, K.~Mönig, M.~Schott, and J.~Stelzer, ``{The global
  electroweak fit at NNLO and prospects for the LHC and ILC},''
  \href{http://dx.doi.org/10.1140/epjc/s10052-014-3046-5}{{\em Eur. Phys. J.}
  {\bfseries C74} (2014) 3046},
\href{http://arxiv.org/abs/1407.3792}{{\ttfamily arXiv:1407.3792 [hep-ph]}}.

\bibitem{Aad:2014cka}
{\bfseries ATLAS} Collaboration, G.~Aad {\em et~al.}, ``{Search for high-mass
  dilepton resonances in pp collisions at $\sqrt{s}=8$??TeV with the ATLAS
  detector},'' \href{http://dx.doi.org/10.1103/PhysRevD.90.052005}{{\em Phys.
  Rev.} {\bfseries D90} no.~5, (2014) 052005},
\href{http://arxiv.org/abs/1405.4123}{{\ttfamily arXiv:1405.4123 [hep-ex]}}.

\bibitem{ATLAS:2014wra}
{\bfseries ATLAS} Collaboration, G.~Aad {\em et~al.}, ``{Search for new
  particles in events with one lepton and missing transverse momentum in $pp$
  collisions at $\sqrt{s}$ = 8 TeV with the ATLAS detector},''
  \href{http://dx.doi.org/10.1007/JHEP09(2014)037}{{\em JHEP} {\bfseries 09}
  (2014) 037},
\href{http://arxiv.org/abs/1407.7494}{{\ttfamily arXiv:1407.7494 [hep-ex]}}.

\bibitem{Khachatryan:2014fba}
{\bfseries CMS} Collaboration, V.~Khachatryan {\em et~al.}, ``{Search for
  physics beyond the standard model in dilepton mass spectra in proton-proton
  collisions at $ \sqrt{s}=8 $ TeV},''
  \href{http://dx.doi.org/10.1007/JHEP04(2015)025}{{\em JHEP} {\bfseries 04}
  (2015) 025},
\href{http://arxiv.org/abs/1412.6302}{{\ttfamily arXiv:1412.6302 [hep-ex]}}.

\bibitem{Khachatryan:2014tva}
{\bfseries CMS} Collaboration, V.~Khachatryan {\em et~al.}, ``{Search for
  physics beyond the standard model in final states with a lepton and missing
  transverse energy in proton-proton collisions at sqrt(s) = 8 TeV},''
  \href{http://dx.doi.org/10.1103/PhysRevD.91.092005}{{\em Phys. Rev.}
  {\bfseries D91} no.~9, (2015) 092005},
\href{http://arxiv.org/abs/1408.2745}{{\ttfamily arXiv:1408.2745 [hep-ex]}}.

\bibitem{Dulat:2015mca}
S.~Dulat, T.~J. Hou, J.~Gao, M.~Guzzi, J.~Huston, P.~Nadolsky, J.~Pumplin,
  C.~Schmidt, D.~Stump, and C.~P. Yuan, ``{The CT14 Global Analysis of Quantum
  Chromodynamics},''
\href{http://arxiv.org/abs/1506.07443}{{\ttfamily arXiv:1506.07443 [hep-ph]}}.

\bibitem{Atlas2WH}
T.~A. collaboration,
``{Search for new resonances decaying to a W or Z boson and a Higgs boson in
  the $\ell\ell b\bar b$, $\ell\nu b\bar b$, and $\nu\nu b\bar b$ channels in
  $pp$ collisions at $\sqrt s = 13$~TeV with the ATLAS detector},''.

\bibitem{CMS-DP-2015-039}
{\bfseries CMS Collaboration} Collaboration, ``{Event Display of a Candidate
  Electron-Positron Pair with an Invariant Mass of 2.9 TeV},''.
  \url{https://cds.cern.ch/record/2048626}.

\bibitem{Hinterbichler:2011tt}
K.~Hinterbichler, ``{Theoretical Aspects of Massive Gravity},''
  \href{http://dx.doi.org/10.1103/RevModPhys.84.671}{{\em Rev. Mod. Phys.}
  {\bfseries 84} (2012) 671--710},
\href{http://arxiv.org/abs/1105.3735}{{\ttfamily arXiv:1105.3735 [hep-th]}}.

\end{thebibliography}\endgroup
\bibliographystyle{utphys}
}

\end{document}